\def\rfr#1{(\ref{#1})}

\def\dert#1#2{\frac{{{d}}{#1}}{{{d}}{#2}}}              % derivate parziali e totali prima e seconda

\def\bb{\bibitem}
\def\eqI{\begin{equation}}
\def\eqF{\end{equation}}
\def\eqIa{\begin{eqnarray}}
\def\eqFa{\end{eqnarray}}
\def\rp#1#2{{#1\over#2}} \def\lb#1{\label{#1}}
\def\virg#1{``#1''}
\def\rf#1{Ref.~\cite{#1}}

\def\bds#1{\boldsymbol{#1}}

\documentclass[11pt]{article}%{ws-ijmpd}
\usepackage{amsmath,amsthm,amscd,amssymb}
\usepackage{latexsym}
\usepackage{graphicx,epsfig}

\begin{document}

\noindent{\bf \LARGE{Classical
and relativistic
orbital motions around a mass-varying body}}
\\
\\
\\
{L. Iorio$^{\ast}$\\
{\it $^{\ast}$INFN-Sezione di Pisa. Address for correspondence: Viale Unit$\grave{a}$ di Italia 68
70125 Bari (BA), Italy.  \\ e-mail: lorenzo.iorio@libero.it}

\vspace{4mm}

\begin{abstract}
 I work out  the Newtonian
 and general relativistic
 effects due to an isotropic mass loss $\dot M/M$ of a body on the orbital motion of a test particle around it; the present analysis is also valid  for a variation $\dot G/G$ of the Newtonian constant of gravitation.
 Concerning the Newtonian case,
 I use the Gauss equations for the variation of the elements and obtain negative secular rates for the osculating semimajor axis $a$, the eccentricity $e$ and the mean anomaly $\mathcal{M}$, while the argument of pericenter $\omega$  does not experience secular precession; the longitude of the ascending node $\Omega$ and the inclination $i$ remain unchanged as well. The anomalistic period is different from  the Keplerian one: it turns out to be larger than it. The true orbit, instead, expands, as shown by a numerical integration of the equations of motion with MATHEMATICA; in fact, this is in agreement with the seemingly counter-intuitive decreasing of $a$ and $e$ because they refer to the osculating Keplerian ellipses which approximate the trajectory at each instant.  A comparison with the results obtained with different approaches by other researchers is made.
 General relativity induces positive secular rates of the semimajor axis  and the eccentricity completely negligible in the present and future evolution of the solar system.
%
%By assuming for the Sun $\dot M/M = -9\times 10^{-14}$  yr$^{-1}$ it turns out that the Earth's perihelion position is displaced outward by 1.3 cm %along the fixed line of apsides  after each revolution. By applying our results  to the phase in which the radius of the Sun, already moved to the Red %Giant Branch of the Hertzsprung-Russell Diagram, will become as large as 1.20 AU in about $1$ Myr, I find that the Earth's perihelion position on the %fixed line of the apsides  will increase by  $\approx 0.22-0.25$ AU (for $\dot M/M = -2\times 10^{-7}$ yr$^{-1}$); other researchers point towards an %increase of $0.37-0.63$ AU. Mercury will be destroyed already at the end of the Main Sequence, while Venus should be engulfed in the initial phase of %the Red Giant Branch phase; the orbits of the outer planets will increase by $1.2-7.5$ AU.  Simultaneous long-term numerical integrations of the %equations of motion of all the major bodies of the solar system, with the inclusion of a mass-loss term in the dynamical force models as well, are %required to check if the mutual N-body interactions may substantially change the picture analytically outlined here, especially in the Red Giant %Branch phase in which Mercury and Venus may be removed from the integration.
 %%Thus, even without invoking tidal effects and drag, the Earth should not avoid the engulfment in the expanded solar photosphere.
\end{abstract}

Keywords gravitation;  stars: mass-loss;   celestial mechanics\\
PACS\ 95.30.Sf; 97.10.Me; 95.10.Ce

\section{Introduction}
In this paper I investigate the classical orbital effects induced by an isotropic variation $\dot M/M$ of the mass of a central body  on the motion of a test particle; my analysis is valid also for a change $\dot G/G$ of the Newtonian constant of gravitation. This problem, although interesting in itself, is not only an academic one because of the relevance that it may have on the ultimate destiny of planetary companions in many stellar systems in which the host star experiences a mass loss, like our Sun \cite{Sch08}.
 With respect to this aspect, my analysis may be helpful  in driving future researches towards the implementation of long-term N-body simulations including the temporal change of $GM$ as well, especially over timescales covering paleoclimate changes, up to
the Red Giant Branch (RGB) phase
in which some of the inner planets should  be engulfed by the expanding Sun.
Another problem, linked to the one investigated here, which has recently received attention is the
observationally determined secular variation of the Astronomical Unit \cite{Kra04,Sta05,Nor08,Kli08}. Moreover, increasing accuracy in astrometry pointing towards microarcsecond level \cite{IAU07}, and
long-term stability in clocks \cite{Osk06} require to consider the possibility that smaller and subtler perturbations will be soon detectable in the solar system.  Also future planetary ephemerides should take into account $\dot M/M$. Other phenomena which may, in principle, show connections with the problem treated here are the secular decrease of the semimajor axes of the LAGEOS satellites, amounting to 1.1 mm d$^{-1}$, \cite{Ruby} and the increase of the lunar orbit's eccentricity of $0.9\times 10^{-11}$ yr$^{-1}$ \cite{luna}.
%However, a detailed analysis of all such issues is beyond the scope of this paper.

Many treatments of the mass loss-driven  orbital dynamics in the framework of the Newtonian mechanics, based on different approaches and laws of variation of the central body's mass, can be found in literature; see, e.g., \cite{Stro,Jea24,Jea29,Arm53,Haj63,Haj66,Khol,Dep83,Kev96,Kra04,Nor08} and references therein. However, they are sometimes rather confused and involved, giving unclear results concerning the behavior of the Keplerian orbital elements and the true orbit.

The plan of the paper is as follows. Section \ref{minchia} is devoted to a theoretical description of the phenomenon in a two-body scenario. By working in the Newtonian framework, I will  analytically work out the changes after one orbital revolution experienced by all the Keplerian orbital elements of a test particle moving in the gravitational field of a central mass experiencing a variation of its $GM$ linear in time. Then, I will clarify the meaning of the results obtained by  performing a numerical integration of the equations of motion in order to visualize the true trajectory followed by the planet.  Concerning the method adopted, I will use the Gauss perturbation equations \cite{Bert,roy}, which are valid for generic disturbing accelerations depending on position, velocity and time,  the \virg{standard} Keplerian orbital elements (the Type I according to, e.g., \rf{Khol}) with the eccentric anomaly $E$ as \virg{fast} angular variable. Other approaches and angular variables like, e.g. the Lagrange perturbation equations \cite{Bert,roy}, the Type II orbital elements \cite{Khol} and the mean anomaly $\mathcal{M}$ could be used, but, in my opinion, at a price of major conceptual and computational difficulties\footnote{Think, e.g.,  about the cumbersome expansions in terms of the mean anomaly and the Hansen coefficients, the subtleties concerning the choice of the independent variable in the Lagrange equations for the semimajor axis and the eccentricity \cite{Bert}.}. With respect to possible connections with realistic situations, it should be noted that, after all, the Type I orbital elements are usually determined or improved in standard data reduction analyses of the motion of planets and (natural and artificial) satellites. Instead, my approach should, hopefully, appear more transparent and easy to interpret, although, at first sight, some counter-intuitive results concerning the semimajor axis and the eccentricity will be obtained; moreover, for the chosen time variation of the mass of the primary, no approximations are used in the calculations which are quite straightforward.  However, it is important to stress that such allegedly puzzling features are only seemingly paradoxical because they will turn out to be in agreement with numerical integrations of the equations of motion, as explicitly shown by the numerous pictures depicted.  Anyway, the interested reader is advised to look also at \rf{Khol} for a different approach.
In Section \ref{due} I will work within the general relativistic gravitoelectromagnetic framework by calculating the gravitoelectric effects on all the Keplerian orbital elements of a freely falling test particle in a non-stationary gravitational field.
%In Section \ref{evol} I will apply our results to the future Sun-Earth scenario and to the other planets of the solar system.
Section \ref{tre} is devoted to a discussion of the findings of other researchers and contains some numerical calculations concerning the previously mentioned orbital phenomena of LAGEOS and the Moon. Section \ref{quattro} summarizes my results.
%
%I wish to make a final remark concerning the field of applicability of our results to  realistic astrophysical contexts. Indeed, throughout the paper %I will consider only a two-body configuration in which the primary undergoes a time-variation of its $GM$. If I want to apply this scenario to the %evolution of the Sun-Earth system over timescales of the order of 0.1-1 Myr and more it should be taken into account that, in principle, also the %other planets  induce relatively large changes in the eccentricity (and the other orbital parameters) of the terrestrial orbit (see \cite{KK} and %references therein; \cite{Lask08}). Simulations looking back in time have shown that this happens on timescales of the order of just 0.1 Myr, and it %even appears to be an important forcing factor for climate changes \cite{Lask04}. Thus, in extending our results to deep-future scenarios,  I might %be wrong, in principle, about how representative the present-day Earth's eccentricity is for any very long timescale (as I will show, the magnitude %of the changes depends on the eccentricity).

%
%
\section{Analytical calculation of the orbital effects by $\dot\mu/\mu$}\lb{minchia}
% In this Section I analytically work out the Newtonian effects of a temporal variation of the $GM$ of the primary on the orbital motion of the %secondary in a two-body scenario.
 %\subsection{The Newtonian scenario}\lb{uno}

By defining
\eqI\mu\equiv GM \eqF at a given epoch $t_0$,
the acceleration of a test particle orbiting a central body experiencing a variation of $\mu$ is, to first order in $t-t_0$,
\eqI \bds{A} =-\rp{\mu(t)}{r^2}\bds{\hat r}\approx -\rp{\mu}{r^2}\left[1 + \left(\rp{\dot\mu}{\mu}\right)(t-t_0)\right]\bds{\hat r},\lb{accel}\eqF
with $\dot\mu\equiv\dot\mu|_{t=t_0}$. $\dot\mu$ will be assumed constant throughout the temporal interval of interest $\Delta t = t-t_0$, as it is, e.g., the case for most of the remaining lifetime of the Sun as a Main Sequence (MS) star \cite{Sch08}.    Note that $\dot\mu$ can, in principle, be due to a variation of both the Newtonian gravitational constant $G$ and the mass $M$ of the central body, so that
\eqI \rp{\dot\mu}{\mu} = \rp{\dot G}{G} + \rp{\dot M}{M}.\eqF
%If $\dot\mu$ is caused by a mass variation of the central body, \rfr{accel} is valid if such a variation is isotropic \cite{Haj63}, as I  will %assume here in view of the application of our results to the real case of the mass-varying Sun \cite{Sch08}.
Moreover, while the orbital angular momentum is conserved, this does not happen for the energy.

By limiting ourselves to realistic astronomical scenarios like our solar system, it is quite realistic to assume that
\eqI \left(\rp{\dot\mu}{\mu}\right)(t-t_0)\ll 1\lb{condiz}\eqF  over most of its remaining lifetime:
 indeed, since  $\dot M/M$ is of the order of\footnote{About $80\%$ of such a mass-loss is due to the core nuclear burning, while the remaining $20\%$ is due to average solar wind.} $10^{-14}$ yr$^{-1}$ for the Sun \cite{Sch08}, the condition \rfr{condiz} is satisfied for the remaining\footnote{The age of the present-day MS Sun is 4.58 Gyr, counted from its zero-age MS star model \cite{Sch08}.} $\approx$ 7.58 Gyr before the Sun will approach the  RGB tip in the Hertzsprung-Russell Diagram (HRD). Thus, I can treat it perturbatively with the standard methods of celestial mechanics.

 The unperturbed  Keplerian ellipse at epoch $t_0$, assumed coinciding with the time of the passage at perihelion $t_p$, is characterized by
\begin{equation}
\begin{array}{lll}

r = a(1-e\cos E),\\\\
dt = \left(\rp{1-e\cos E}{n}\right)dE,\\\\
\cos f = \rp{\cos E - e}{1-e\cos E},\\\\
\sin f = \rp{\sqrt{1-e^2}\sin E}{1-e\cos E},
\end{array}\lb{cofi}
 \end{equation}
where $a$ and $e$ are the semimajor axis and the eccentricity, respectively, which fix the size and the shape of the unchanging Keplerian orbit, $n=\sqrt{\mu/a^3}$ is its unperturbed Keplerian mean motion, $f$ is the true anomaly, reckoned from the pericentre, and $E$ is the eccentric anomaly.
This would be the path followed by the particle for any $t>t_p$ if the mass loss would suddenly cease at $t_p$. Instead, the true path will be different because of the perturbation induced by $\dot\mu$ and the orbital parameters of the osculating ellipses approximating the real trajectory at each instant of time will slowly change in time.

\subsection{The semimajor axis and the eccentricity}
The Gauss equation for the variation of the semimajor axis $a$ is   \cite{Bert,roy}
\eqI\dert a t = \rp{2}{n\sqrt{1-e^2}} \left[e A_r\sin f +A_{\tau}\left(\rp{p}{r}\right)\right],\lb{gaus_a}\eqF
where $A_{r}$  and $A_{\tau}$ are the radial and transverse, i.e. orthogonal to the direction of $\bds{\hat r}$, components, respectively, of the disturbing acceleration, and $p=a(1-e^2)$ is the semilatus rectum.
In the present case
\eqI A\equiv A_r = -\rp{\dot\mu}{r^2}(t-t_p),\lb{dist}\eqF i.e. there is an entirely radial perturbing acceleration. For $\dot\mu<0,$ i.e. a decrease in the body's $GM$, the total gravitational attraction felt by the test particle, given by \rfr{accel}, is reduced with respect to the epoch $t_p$.
In order to have the rate of the semimajor axis averaged over one (Keplerian) orbital revolution \rfr{dist} must be inserted into \rfr{gaus_a},   evaluated onto the unperturbed Keplerian ellipse with \rfr{cofi} and finally integrated over $ndt/2\pi$ from 0 to $2\pi$ because $n/2\pi=1/P^{\rm Kep}$ (see below).
Note that, from \rfr{cofi}, it can be obtained
\eqI t-t_p = \rp{E -e\sin E}{n}.\eqF
As a result, I have
\eqI\left\langle\dert a t\right\rangle  =  -\rp{e}{\pi}\left(\rp{\dot\mu}{\mu}\right)a
\int_0^{2\pi} \rp{\left(E -e\sin E\right)\sin E}{\left(1-e\cos E\right)^2} dE = 2\left(\rp{e}{1-e}\right)\left(\rp{\dot\mu}{\mu}\right)a.\lb{arate}
\eqF
Note that if $\mu$ decreases $a$ gets reduced as well: $\left\langle\dot a\right\rangle< 0$. This may be seemingly bizarre and counter-intuitive, but, as it will be shown  later, it is not in contrast with the true orbital motion.

The Gauss equation for the variation of the eccentricity is  \cite{Bert,roy}
\eqI \dert e t  = \rp{\sqrt{1-e^2}}{na}\left\{A_r\sin f + A_{\tau}\left[\cos f + \rp{1}{e}\left(1 - \rp{r}{a}\right)\right]\right\}.\lb{gaus_e}\eqF
For $A=A_r$, it reduces to
\eqI\dert e t = \left(\rp{1-e^2}{2ae}\right)\dert a t,\eqF
so that
\eqI\left\langle\dert e t\right\rangle = (1+e)\left(\rp{\dot\mu}{\mu}\right);\lb{erate}\eqF
also the eccentricity gets smaller for $\dot\mu<0$.

 As a consequence of the found variations of the osculating semimajor axis and the eccentricity, the osculating orbital angular momentum per unit mass, defined by $L^2 = \mu a(1-e^2)$, remains constant: indeed, by using \rfr{arate} and \rfr{erate}, it turns out
\eqI \left\langle\rp{d L^2}{dt}\right\rangle = \mu\left\langle\dot a\right\rangle(1-e^2)-2\mu a e \left\langle\dot e\right\rangle = 0.\lb{momang}\eqF

The osculating total energy $\mathcal{E}=-\mu/2a$ decreases according to
\eqI\left\langle\dert {\mathcal{E}} t\right\rangle = \rp{\mu}{2 a^2}\left\langle\dot a\right\rangle = \left(\rp{e}{1-e}\right)\rp{\dot\mu}{a}.\lb{enosc}\eqF

Moreover, the osculating Keplerian period \eqI P^{\rm Kep}\equiv \rp{2\pi}{n}=2\pi\sqrt{\rp{a^3}{\mu}},\eqF which, by definition, yields the time elapsed between two consecutive perihelion crossings in absence of perturbation, i.e. it is the time required to describe a fixed osculating Keplerian ellipse, decreases according to
\eqI\left\langle\dert {P^{\rm Kep}} t\right\rangle = \rp{3}{2}P^{\rm Kep}\rp{\left\langle\dot a\right\rangle}{a}=\rp{6\pi e\dot\mu}{(1-e)}\left(\rp{a}{\mu}\right)^{3/2}.\lb{pke}\eqF As I will show, also such a result is not in contrast with the genuine orbital evolution.

\subsection{The pericentre, the node and the inclination}
The Gauss equation for the variation of the pericentre $\omega$ is  \cite{Bert,roy}
\eqI
\dert\omega t  = \rp{\sqrt{1-e^2}}{nae}\left[-A_r\cos f + A_{\tau}\left(1+\rp{r}{p}\right)\sin f\right]-\cos i\dert\Omega t,\lb{gaus_o}
\eqF
where $i$ and $\Omega$ are the the inclination and the longitude of the ascending node, respectively, which fix the orientation of the osculating ellipse in the inertial space.
Since $d\Omega/dt$ and $di/dt$ depend on the normal component $A_{\nu}$ of the disturbing acceleration, which is absent in the present case, and $A=A_r$, I have
\eqI
\left\langle\dert \omega t\right\rangle =\rp{\sqrt{1-e^2}}{2\pi e}\left(\rp{\dot\mu}{\mu}\right)\int_0^{2\pi}\rp{(E-e\sin E)(\cos E- e)}{(1-e\cos E)^2}dE=0:
\eqF
the osculating ellipse does not change its orientation in the orbital plane, which, incidentally, remains fixed in the inertial space because $A_{\nu}=0$ and, thus, $d\Omega/dt = di/dt = 0$.

\subsection{The mean anomaly}
The Gauss equation for the mean anomaly $\mathcal{M}$, defined as $\mathcal{M} = n(t-t_p)$,   \cite{Bert,roy}
is
\eqI\dert {\mathcal{M}} t = n - \rp{2}{na} A_r\rp{r}{a} -\sqrt{1-e^2}\left(\dert\omega t + \cos i \dert\Omega t\right).\lb{gaus_M}\eqF
It turns out that, since
\eqI -\rp{2}{na}A_r\rp{r}{a}dt = \rp{2\dot\mu}{n^3 a^3}(E-e\sin E)dE,\eqF
\eqI \left\langle\dert {\mathcal{M}} t\right\rangle = n + 2\pi\left(\rp{\dot\mu}{\mu}\right);\eqF the mean anomaly changes uniformly in time at a slower rate with respect to the unperturbed Keplerian case for $\dot\mu< 0$.
\subsection{Numerical integration of the equations of motion and explanation of the seeming contradiction with the analytical results}
At first sight, the results obtained here may be rather confusing: if the gravitational attraction of the Sun reduces in time because of its mass loss the orbits of the planets should expand (see the trajectory plotted in Figure \ref{picture}, numerically integrated with MATHEMATICA), while I obtained that the semimajor axis and the eccentricity undergo secular decrements.
\begin{figure}
\includegraphics[width=\columnwidth]{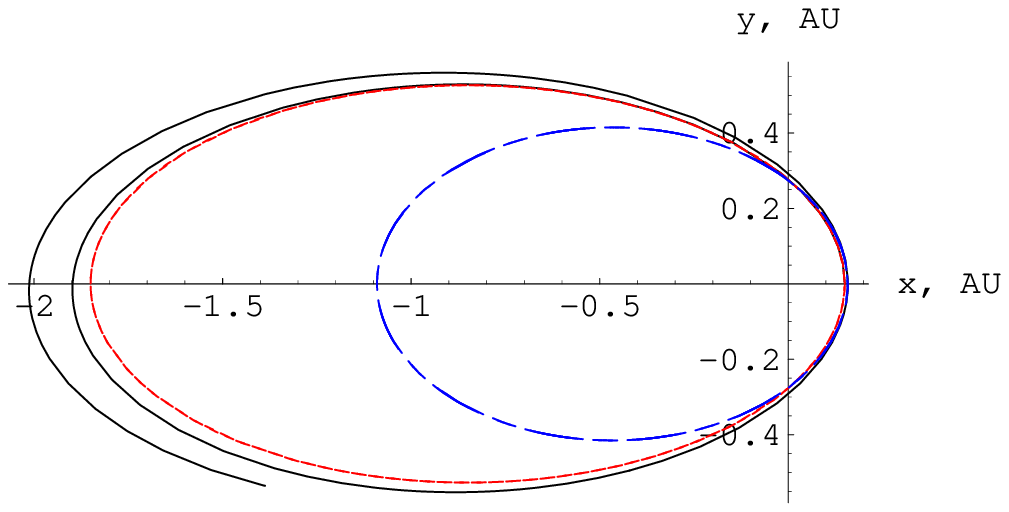}
 \caption{\label{picture} Black continuous line: true trajectory obtained by numerically integrating with MATHEMATICA the perturbed equations of motion in Cartesian  coordinates over 2 yr; the disturbing acceleration \rfr{accel} has been adopted. The planet starts from the perihelion on the \textsc{x} axis. Just for illustrative purposes, a mass loss rate of the order of $10^{-2}$ yr$^{-1}$ has been adopted for the Sun; for the planet initial conditions corresponding to $a=1$ AU, $e=0.8$ have been chosen. Red dashed line: unperturbed Keplerian ellipse at $t = t_0= t_p$. Blue dash-dotted line: osculating Keplerian ellipse after the first perihelion passage. As can be noted, its semimajor axis and eccentricity are clearly smaller than those of the initial unperturbed ellipse. Note also that after 2 yr  the planet has not yet reached the perihelion as it would have done in absence of mass loss, i.e. the true orbital period is longer than the Keplerian one of the osculating red ellipse.}
\end{figure}
Moreover, I found that the Keplerian period $P^{\rm Kep}$
%, which does not coincide with the Keplerian one,
decreases,  while one would expect that the orbital period increases.

In fact, there is no contradiction, and my analytical results do yield us realistic information on the true evolution of the planetary motion. Indeed, $a$, $e$ and $P^{\rm Kep}$ refer to the osculating Keplerian ellipses which, at any instant, approximate the true trajectory; it, instead, is not an ellipse, not being  bounded. Let us start at $t_p$ from the osculating pericentre of the Keplerian ellipse corresponding to chosen initial conditions: let us use a heliocentric frame with the  \textsc{x} axis oriented along the osculating pericentre. After a true revolution, i.e. when the true radius vector of the planet has swept an angular interval of $2\pi$, the planet finds itself again on the  \textsc{x} axis, but at a larger distance from the starting point because of the orbit expansion induced by the Sun's mass loss. It is not difficult to understand that the osculating Keplerian ellipse approximating the trajectory at this perihelion passage is oriented as before because there is no variation of the (osculating) argument of pericentre,  but has smaller semimajor axis and eccentricity.
%; thus, it does not surprise that the related osculating anomalistic period decreases.
And so on, revolution after revolution, until the perturbation theory can be applied, i.e. until $\dot\mu/\mu (t-t_p)<<1$.  In Figure \ref{picture} the situation described so far is qualitatively illustrated. Just for illustrative purposes I enhanced the overall effect by assuming  $\dot\mu/\mu \approx 10^{-2}$ yr$^{-1}$ for the Sun; the initial conditions for the planet correspond to an unperturbed Keplerian ellipse with $a=1$ AU, $e=0.8$ with the present-day value of the Sun's mass in one of its foci. It is apparent that the initial osculating red dashed ellipse has larger $a$ and $e$ with respect to the second osculating blue dash-dotted ellipse.
Note also that the true orbital period, intended as the time elapsed between two consecutive crossings of the perihelion, is larger than the  unperturbed Keplerian one of the initial red dashed osculating ellipse, which would amount to 1 yr for the Earth: indeed, after 2 yr the planet has not yet reached the perihelion for its second passage.
%In the following I will obtain such a result analytically.

Now, if I compute the radial change $\Delta r(E)$ in the osculating radius vector as a function of the eccentric anomaly $E$ I can gain useful insights concerning how much the true path has expanded after two consecutive perihelion passages.
From the Keplerian expression of the Sun-planet distance
\eqI r = a(1-e\cos E)\eqF one gets the radial component of the orbital perturbation  expressed in terms of the eccentric anomaly $E$
\eqI \Delta r (E) = (1-e\cos E)\ \Delta a-a\cos E\ \Delta e + ae\sin E\ \Delta E;\eqF
it agrees with the results obtained by, e.g., Casotto in \rf{Cas93}.
Since
%%%%%%%%%%%%%%%%%%%%%%%%%%%%%%%%%%%%%
\eqI
\left\{
\begin{array}{lll}
\Delta a & = & -\rp{2ae}{n}\left(\rp{\dot\mu}{\mu}\right)\left(\rp{\sin E - E\cos E}{1-e\cos E}\right),\\\\
\Delta e & = & -\rp{(1-e^2)}{n}\left(\rp{\dot\mu}{\mu}\right)\left(\rp{\sin E - E\cos E}{1-e\cos E}\right),\lb{ecces}\\\\
\Delta E & = & \left(\rp{\Delta {\mathcal{M}} +\sin E\ \Delta e  }{1-e\cos E}\right)=\rp{1}{n}\left(\rp{\dot\mu}{\mu}\right)\left[\mathcal{A}(E)+\mathcal{B}(E)+\mathcal{C}(E)\right],
 \end{array}
\right.
\eqF
%%%%%%%%%%%%%%%%%%%%%%%%%%%%%%%%%%%%%%%
with
%%%%%%%%%%%%%%%%%%%%%%%%%%%%%%%%%
\eqI
\left\{
 \begin{array}{lll}
\mathcal{A}(E) & = & \rp{E^2 + 2e(\cos E -1)}{1-e\cos E},\\\\
\mathcal{B}(E) & = & \left(\rp{1-e^2}{e}\right)\left[\rp{1+e - (1+e)\cos E - E\sin E}{(1-e\cos E)^2}\right],\\\\
\mathcal{C}(E) & = & -\rp{(1-e^2)\sin E(\sin E - e\cos E)}{(1-e\cos E)^2},
\end{array}
\right.
\eqF
%%%%%%%%%%%%%%%%%%%%%%%%%%%%%%%%%%%%%%%%%%%%%%%%%%%%%%%
%
%
%
it follows
\eqI \Delta r(E) = \rp{a}{n}\left(\rp{\dot\mu}{\mu}\right)\left[\mathcal{D}(E)+\mathcal{F}(E)\right], \lb{mega}\eqF
with
%%%%%%%%%%%%%%%%%%%%%%%%%%%%%%%%%
\eqI
\left\{
\begin{array}{lll}
\mathcal{D}(E) & = & e\left[-2(\sin E-E \cos E) + \rp{\sin E\left[E^2 + 2e(\cos E -1)\right]}{1-e\cos E} -\rp{(1-e^2)\sin^2 E(\sin E-e\cos E)}{(1-e\cos E)^2} \right],\\\\
\mathcal{F}(E) & = & \left(\rp{1-e^2}{1-e\cos E}\right)\left\{
\cos E(\sin E - E \cos E) + \sin E\left[\rp{1+e - (1+e)\cos E-E \sin E}{1-e\cos E}\right]
\right\}.
\end{array}
\right.
\lb{mega2}\eqF
%%%%%%%%%%%%%%%%%%%%%%%%%%%%%%%%%%%%%%%%%%%%%%%%%%%%%%%
From \rfr{mega} and \rfr{mega2} it turns out that for $E>0$ $\Delta r(E)$ never vanishes; after one  orbital revolution, i.e. after that an angular interval of $2\pi$ has been swept by the (osculating) radius vector, a net increase of the radial (osculating) distance occurs according to\footnote{According to \rfr{mega} and \rfr{mega2}, $\Delta r (0)=0$.} \eqI\Delta r (2\pi) - \Delta r (0)= \Delta r (2\pi)  = -\rp{2\pi}{n}a\left(\rp{\dot\mu}{\mu}\right)(1-e).\lb{deltaerre}\eqF
This analytical result is qualitatively confirmed by the difference\footnote{Strictly speaking, $\Delta r$ and the quantity plotted in Figure \ref{deltaerref} are different objects, but, as the following discussion will clarify, I can assume that, in practice, they are the same.} $\Delta r(t)$ between the radial distances obtained from the solutions of two numerical integrations  with MATHEMATICA of the equations of motion over 3 yr with and without $\dot\mu/\mu$; the initial conditions are the same. For illustrative purposes I used $a=1$ AU, $e=0.01$, $\dot\mu/\mu=-0.1$ yr$^{-1}$. The result is depicted in Figure \ref{deltaerref}.
\begin{figure}
\includegraphics[width=\columnwidth]{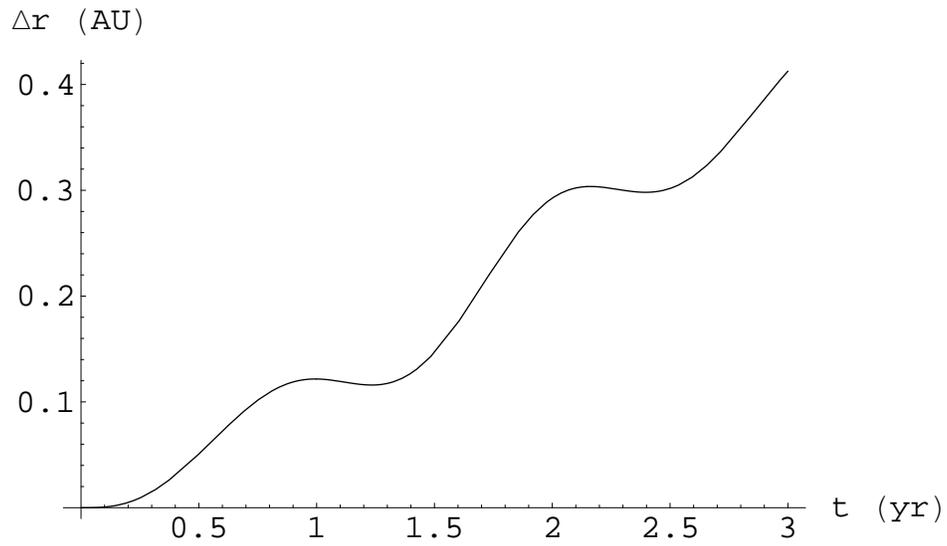}
\caption{\label{deltaerref}  Difference $\Delta r(t)$ between the radial distances obtained from the solutions of two numerical integrations  with MATHEMATICA of the equations of motion over 3 yr  with and without $\dot\mu/\mu$; the initial conditions are the same. Just for illustrative purposes a mass loss rate of the order of $-10^{-1}$ yr$^{-1}$ has been adopted for the Sun; for the planet initial conditions corresponding to $a=1$ AU, $e=0.01$ have been chosen. The cumulative increase of the Sun-planet distance induced by the mass loss is apparent.}
\end{figure}
Note also that  \rfr{mega} and \rfr{mega2} tell us that the shift at the aphelion is
\eqI\Delta r(\pi)=\rp{1}{2}\left(\rp{1+e}{1-e}\right)\Delta r(2\pi),\lb{aphe}\eqF in agreement with Figure \ref{picture} where it is 4.5 times larger than the shift at the perihelion.

Since Figure \ref{picture} tells us that the orbital period gets larger than the Keplerian one, it means that the true orbit must somehow remain behind with respect to the Keplerian one. Thus, a negative perturbation $\Delta\tau$ in the transverse direction must occur as well; see Figure \ref{AUD}.
\begin{figure}
\includegraphics[width=\columnwidth]{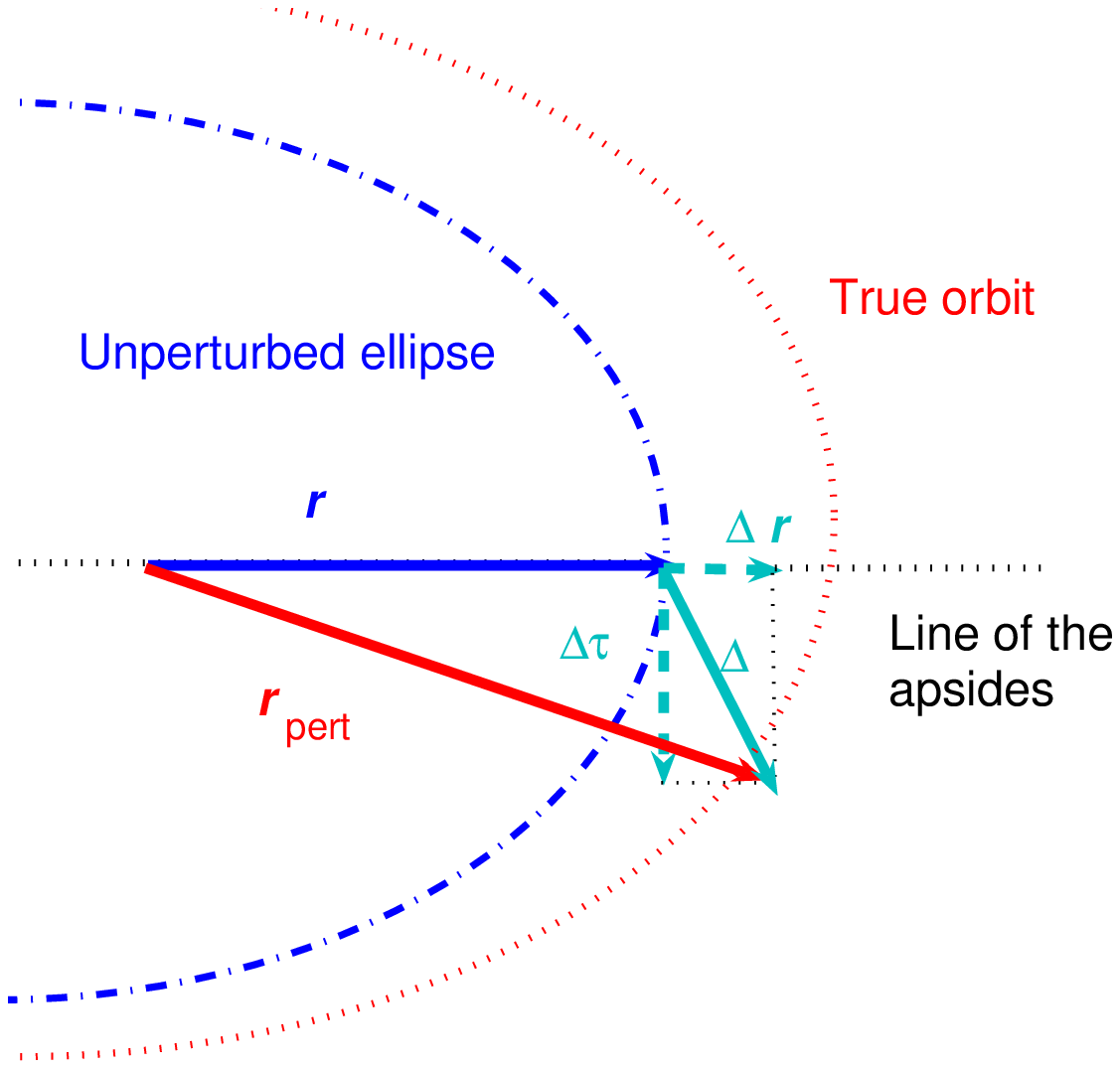}
\caption{\label{AUD}  Radial and transverse perturbations $\bds{\Delta r}$ and $\bds{\Delta \tau}$ of the Keplerian radius vector (in blue); the presence of the transverse perturbation $\bds{\Delta \tau}$  makes the real orbit (in red) lagging behind the Keplerian one.}
\end{figure}

Let us now analytically compute it.
According to \rf{Cas93}, it can be used
\eqI\Delta\tau = \rp{a\sin E}{\sqrt{1-e^2}} + a\sqrt{1-e^2}\ \Delta E + r(\Delta\omega+\Delta\Omega\cos i).\lb{caso}\eqF
By recalling that, in the present case, $\Delta\Omega=0$ and using
\eqI\Delta\omega= -\rp{\sqrt{1-e^2}}{ne}\left(\rp{\dot\mu}{\mu}\right)\left[\rp{1 + e - (1 + e)\cos E - E \sin E}{1-e\cos E}\right],\lb{peris}\eqF
it is possible to obtain from \rfr{ecces} and \rfr{peris}
\eqI\Delta\tau(E) = \rp{a}{n}\left(\rp{\dot\mu}{\mu}\right)\rp{\sqrt{1-e^2}}{\left(1-e\cos E\right)}\left[\mathcal{G}(E) + \mathcal{H}(E)+\mathcal{I}(E)+\mathcal{J}(E)+\mathcal{K}(E)\right],\lb{tra}\eqF
with
%%%%%%%%%%%%%%%%%%%%%%%%%%%%%%%%%
\eqI
\left\{
 \begin{array}{lll}
\mathcal{G}(E) & = & \sin E(E\cos E -\sin E),\\\\
\mathcal{H}(E) & = & \rp{(1-e\cos E)}{e}\left[(1+e)(\cos E-1) + E\sin E\right],\lb{vuuu}\\\\
\mathcal{I}(E) & = & E^2 + 2 e (\cos E-1),\\\\
\mathcal{J}(E) & = &\sin E\left[\rp{(1-e^2)(e\cos E-\sin E)}{1-e\cos E}\right]\\\\
\mathcal{K}(E) & = & \left(\rp{1-e^2}{e}\right)\left[\rp{(1+e)(1-e\cos E) - E\sin E}{1-e\cos E}\right].
\end{array}
\right.
\eqF
%%%%%%%%%%%%%%%%%%%%%%%%%%%%%%%%%%%%%%%%%%%%%%%%%%%%%%%
%
%
%
%
From \rfr{tra} and \rfr{vuuu}
it turns out that for $E>0$ $\Delta \tau(E)$ never vanishes; at the (osculating) time of perihelion passage
\eqI\Delta\tau(2\pi) -\Delta\tau(0)=\rp{4\pi^2}{n}a\left(\rp{\dot\mu}{\mu}\right)\sqrt{\rp{1+e}{1-e}}<0.\eqF
This means that when the Keplerian path has reached the perihelion, the perturbed orbit is still behind it.
%This lag increases by a factor 4 at the perihelion.
 Such features are qualitatively confirmed by Figure \ref{picture}.

From a vectorial point of view, the radial and transverse perturbations to the Keplerian radius vector $\bds r$ yield a correction
\eqI \bds\Delta = \Delta r\ \bds{\hat{r}} + \Delta\tau\ \bds{\hat{\tau}},\eqF so that
\eqI \bds r_{\rm pert} = \bds r + \bds\Delta.\eqF The length of $\bds\Delta$ is
\eqI\Delta(E) =\sqrt{\Delta r(E)^2 + \Delta \tau(E)^2};\eqF \rfr{deltaerre} and \rfr{tra} tell us that at perihelion it amounts to \eqI\Delta(2\pi)=\Delta r(2\pi)\sqrt{1 + {4\pi^2}\rp{(1+e)}{(1-e)^3}}.\eqF
The angle $\xi$ between $\bds\Delta$ and $\bds r$ is given by
\eqI\tan\xi(E) = \rp{\Delta \tau(E)}{\Delta r(E)};\eqF at perihelion it is  \eqI\tan\xi(2\pi)=-2\pi\rp{\sqrt{1+e}}{(1-e)^{3/2}},\eqF i.e. $\xi$ is close to $-90$ deg; for the Earth it is $-81.1$ deg.
Thus, the difference $\delta $ between the lengths of the perturbed radius vector $r_{\rm pert}$ and the Keplerian one $r$ at a given instant amounts to about
\eqI \delta\approx\Delta\cos\xi;\eqF  if fact, this is precisely the quantity determined over 3 yr by the numerical integration of Figure \ref{deltaerref}. At the perihelion I have
\eqI \delta = \Delta r(2\pi)\sqrt{1 + {4\pi^2}\rp{(1+e)}{(1-e)^3}}\cos\xi;\eqF
since for the Earth \eqI\sqrt{1 + {4\pi^2}\rp{(1+e)}{(1-e)^3}}\cos\xi=1.0037,\eqF
it holds \eqI\delta\approx\Delta r(2\pi).\eqF This explains why Figure \ref{deltaerref} gives us just $\Delta r$.

Since the approximate calculations of other researchers often refer to circular orbits, and in view of the fact that when a Sun-like star evolves into a giant tidal interactions circularize\footnote{This fact has been quantitatively proven by the observation of convective binary stars \cite{Beech}.} the orbit of a planet \cite{Zahn}, it is interesting to consider also such limiting case in which other nonsingular osculating orbital elements must be adopted. The eccentricity and the pericentre lose their meaning: thus, it is not surprising that \rfr{erate}, although formally valid for $e\rightarrow 0$, yields a meaningless result, i.e. the eccentricity would become negative. Instead, the semimajor axis is still valid and \rfr{arate} predicts that $\left\langle\dot a\right\rangle=0$ for  $e\rightarrow 0$. The constancy of the osculating semimajor axis is not in contrast with the true trajectory, as clearly showed by Figure \ref{figDue}.
\begin{figure}
\includegraphics[width=\columnwidth]{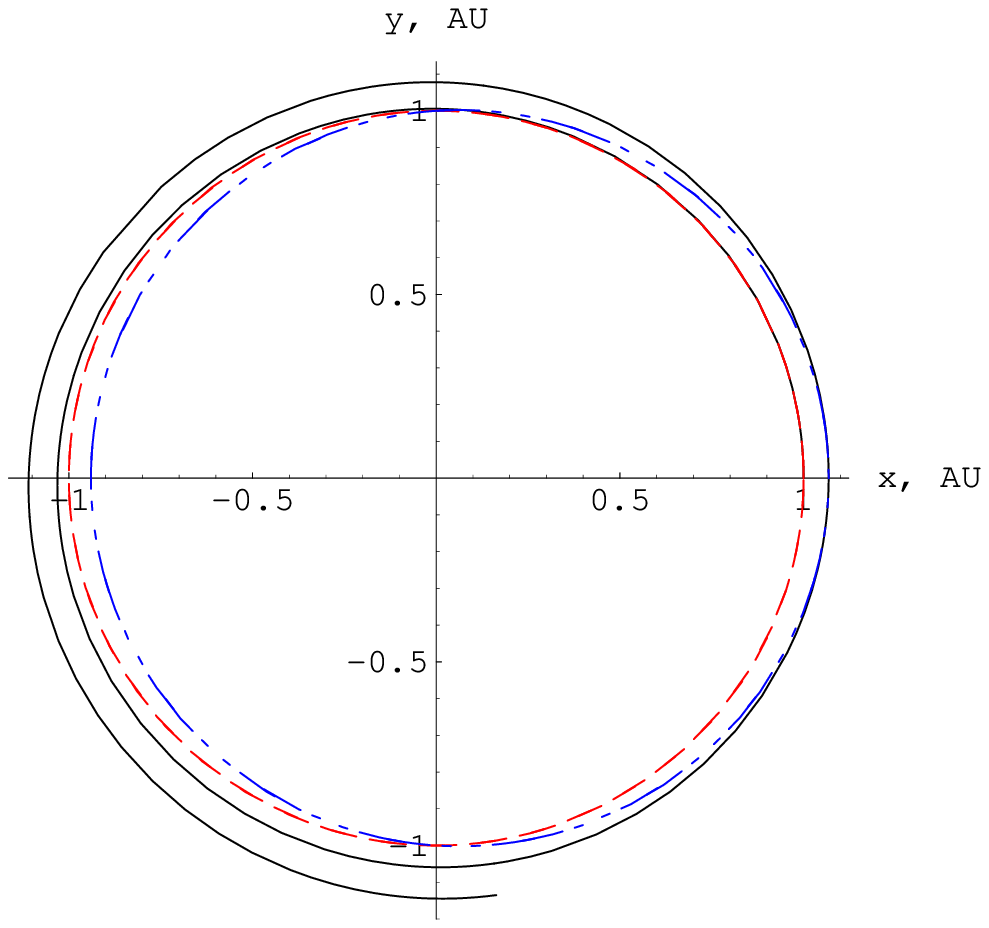}
\caption{\label{figDue} Black continuous line: true trajectory obtained by numerically integrating with MATHEMATICA the perturbed equations of motion in Cartesian  coordinates over 2 yr ; the disturbing acceleration \rfr{accel} has been adopted. The planet starts from a point on the \textsc{x} axis. Just for illustrative purposes, a mass loss rate of the order of $-10^{-2}$ yr$^{-1}$ has been adopted for the Sun; for the planet initial conditions corresponding to $a=1$ AU, $e=0.0$ have been chosen. Red dashed line: unperturbed Keplerian circle at $t = t_0$. Blue dash-dotted line: osculating Keplerian circle after the first  \textsc{x} axis crossing. As can be noted, its semimajor axis and eccentricity are equal to those of the initial unperturbed circle. Note also that after 2 yr  the planet has not yet reached the  \textsc{x} axis as it would have done in absence of mass loss.}
\end{figure}
Again, the true orbital period is larger than the Keplerian one which,  contrary to the eccentric case, remains fixed.
Since $\mathcal{D}(E)=0$ for $e=0$ and $\left.\mathcal{F}(2\pi)\right|_{e=0}=-2\pi,\ \left.\mathcal{F}(0)\right|_{e=0}=0$,
the radial shift per revolution is
\eqI\left.\Delta r(2\pi)\right|_{e=0} = -\rp{2\pi}{n}a\left(\rp{\dot\mu}{\mu}\right).\lb{circul}\eqF
Also in this case the secular increase of the radial distance is present, as qualitatively shown by Figure \ref{deltaerreo}.
%\Rfr{aphe} tell us that, in this case,
%\eqI\Delta r(\pi)=\rp{\Delta r(2\pi)}{2},\eqF in agreement with Figure \ref{figDue}.
Concerning $\Delta\tau$,
after $2\pi$ it is
\eqI\Delta\tau(2\pi)=\rp{4\pi^2}{n}a\left(\rp{\dot\mu}{\mu}\right);\eqF
also in this case, the orbital period is larger than the unperturbed one.
\begin{figure}
\includegraphics[width=\columnwidth]{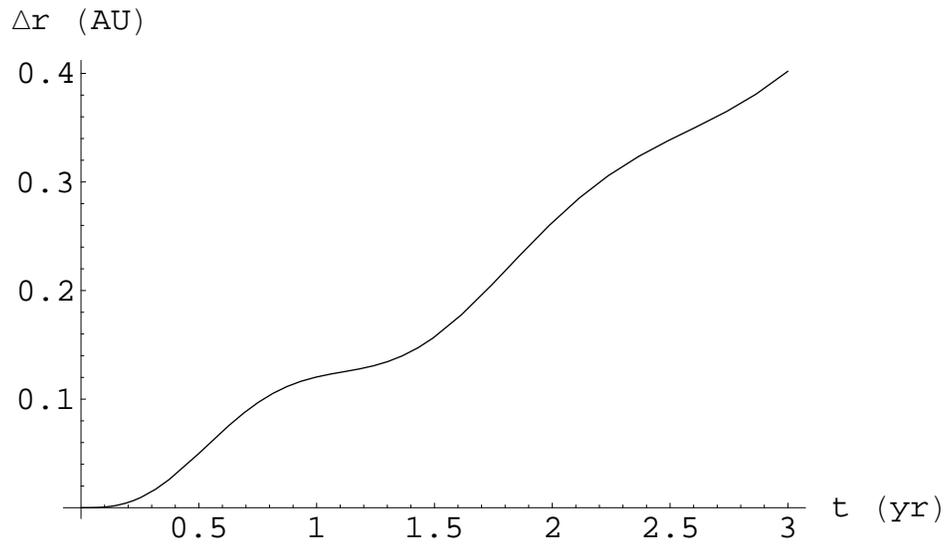}
\caption{\label{deltaerreo}  Difference $\Delta r(t)$ between the radial distances obtained from the solutions of two numerical integrations with MATHEMATICA of the equations of motion over 3 yr with and without $\dot\mu/\mu$; the initial conditions are the same. Just for illustrative purposes a mass loss rate of the order of $10^{-1}$ yr$^{-1}$ has been adopted for the Sun; for the planet initial conditions corresponding to $a=1$ AU, $e=0.0$ have been chosen. The cumulative increase of  the Sun-planet distance induced by the mass loss is apparent.}
\end{figure}

\section{The general relativistic case}\lb{due}
The field equations of general relativity are non-linear, but in the slow-motion $(\bds{\beta}=\bds{v}/c\ll 1)$ and weak-field ($U/c^2\ll 1$) approximation   they get linearized resembling to the linear equations of the Maxwellian electromagnetism; here $v$ and $U$ are the magnitudes of the typical velocities and the gravitational potential of the problem under consideration. This scenario is known as gravitoelectromagnetism \cite{Mash01,Mash07}. In this case the space-time metric is given by
\eqI
ds^2  =  \left(1-2\rp{\Phi}{c^2}\right)c^2dt^2 +\rp{4}{c}\left(\bds{H}\cdot \bds{dr}\right)dt-\left(1+2\rp{\Phi}{c^2}\right)\delta_{ij}dx^i dx^j,
\eqF
where, far from the source, the dominant contributions to the gravitoelectromagnetic potentials can be expressed as
\eqI \Phi = \rp{\mu}{r},\ \bds{H} = \rp{G}{c}\rp{\bds{J}\times \bds{r}}{r^3}.\eqF Here $\bds{J}$ is the proper angular momentum of the central body of mass $M$ and $r$ is so that $r\gg GM/c^2$ and $r\gg J/(Mc)$; $c$ is the speed of light in vacuum.
For a non-stationary source the geodesic equations of motion yield \cite{Bin08}, among other terms, $-\beta^i(3-\beta^2)\Phi_{,0}, i=1,2,3$ which, to order $\mathcal{O}(c^{-2})$, reduces to
\eqI \bds{A} = -3\rp{\dot\mu}{c^2}\rp{\bds{v}}{r}=-3\rp{GM}{c^2}\left(\rp{\dot G}{G} + \rp{\dot M}{M}\right)\rp{\bds{v}}{r}\lb{GEM}.\eqF
 For $\dot\mu<0$ such a perturbing acceleration is directed along the velocity of the test particle.
Although of no practical interest, being of the order of $10^{-24}$ m s$^{-2}$ in the case of a typical Sun-planet system with $\dot M/ M = -9\times 10^{-14}$ yr$^{-1}$, I will explicitly work out the orbital effects of \rfr{GEM};
the effects of the temporal variations of $\bds{J}$ have already been worked out elsewhere \cite{Ior02,Bin08}.
Also in this case I will use the Gauss perturbative case.
Since the radial and transverse components of the unperturbed velocity are
\eqI v_r = \rp{nae\sin f}{\sqrt{1-e^2}},\eqF
\eqI v_{\tau} = \rp{na(1 + e\cos f)}{\sqrt{1-e^2}},\eqF
the radial and transverse components of \rfr{GEM}, evaluated onto the unperturbed Keplerian orbit, are
\eqI A_r = -\rp{3\dot\mu}{c^2} \rp{n e\sin E} {(1-e\cos E)^2},\eqF
\eqI A_{\tau} = -\rp{3\dot\mu}{c^2}\rp{ n \sqrt{1-e^2}} {(1-e\cos E)^2}.\eqF
After lengthy calculations they yield
\eqI\left\langle\dert a t\right\rangle = -\rp{6\dot\mu}{c^2}\left(\rp{2}{\sqrt{1-e^2}}-1\right),\eqF
\eqI\left\langle\dert e t\right\rangle = \rp{3\dot\mu\mu}{c^2 a^4 e}(2-e)\left(1-\rp{1}{\sqrt{1-e^2}}\right).\eqF
Contrary to the classical case, now both the osculating semimajor axis and the eccentricity increase for $\dot\mu<0$.
It turns out that the pericentre and the mean anomaly do not secularly precess. Also in this case the inclination and the node are not affected %because A_{\nu}=0$.
The qualitative features of the motion with the perturbation \rfr{GEM} are depicted in Figure \ref{GR} in which the magnitude of the relativistic term has been greatly enhanced for illustrative purposes.
\begin{figure}
\includegraphics[width=\columnwidth]{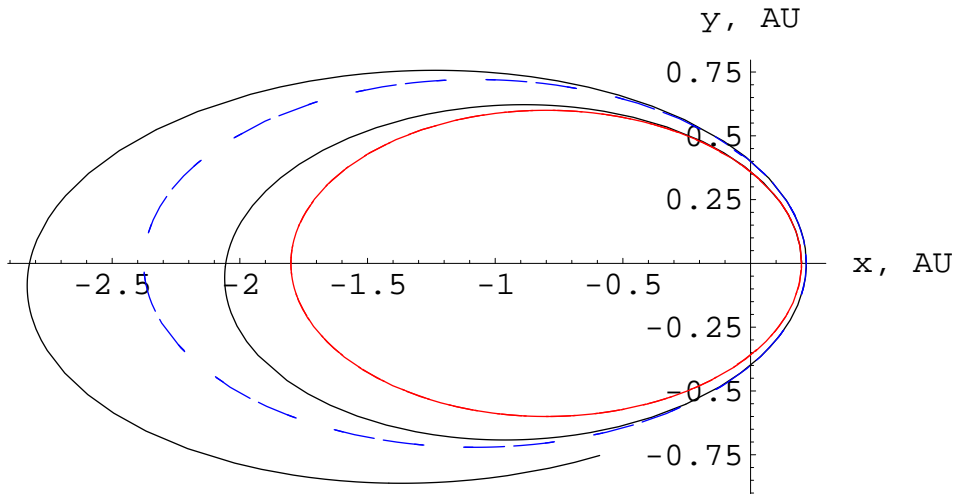}
 \caption{\label{GR} Black continuous line: true trajectory obtained by numerically integrating over 3 yr   the equations of motion perturbed by
 \rfr{GEM}. The planet starts from the perihelion on the $x$ axis. Just for illustrative purposes, a factor $-3\dot\mu/c^2$ of the order of $5\times
 10^{-2}$ AU yr$^{-1}$ has been adopted for the Sun; for the planet initial conditions corresponding to $a=1$ AU, $e=0.8$ have been chosen. Red dashed
 line: unperturbed Keplerian ellipse at $t = t_0= t_p$. Blue dash-dotted line: osculating Keplerian ellipse after the first perihelion passage. As can
 be noted, its semimajor axis and eccentricity are larger than those of the initial unperturbed ellipse.}
\end{figure}

\section{Discussion of other approaches and numerical calculations}\lb{tre}
Here I will briefly review some of the results obtained by others by comparing with ours.

Hadjidemetriou in \rf{Haj63} uses a tangential perturbing acceleration proportional to the test particle's velocity $\bds{v}$,
\eqI \bds{A} = -\rp{1}{2}\left[\rp{\dot\mu}{\mu(t)}\right]\bds{v},\lb{haj}\eqF and a different perturbative approach by finding that, for a generic mass loss, the semimajor axis secularly increases and the eccentricity remains constant. In fact, with the approach followed here it would be possible to show that, to first order in $(\dot\mu/\mu)(t-t_0)$, $\left\langle\dot a\right\rangle = -(\dot\mu/\mu)a$ and $\left\langle\dot e\right\rangle=0$ and that the true orbit is expanding, although in a different way with respect to \rfr{accel} as depicted by Figure \ref{Hadj} in which the magnitude of the mass-loss has been exaggerated for better showing its orbital effects.
\begin{figure}
\includegraphics[width=\columnwidth]{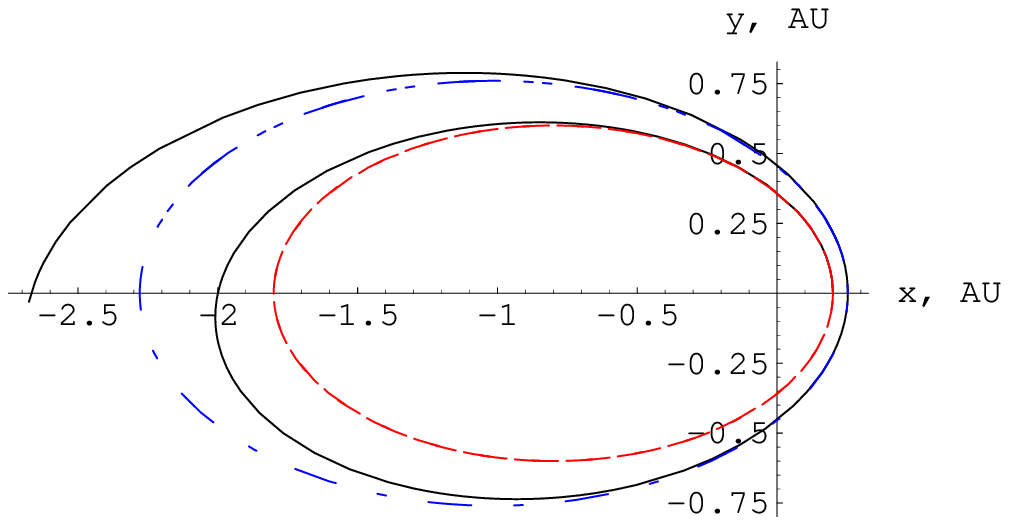}
 \caption{\label{Hadj} Black continuous line: true trajectory obtained by numerically integrating with MATHEMATICA the perturbed equations of motion in Cartesian  coordinates over 2 yr ; the disturbing acceleration \rfr{haj} has been used. The planet starts from the perihelion on the  \textsc{x} axis. Just for illustrative purposes, a mass loss rate of the order of $-10^{-1}$ yr$^{-1}$ has been adopted for the Sun; for the planet initial conditions corresponding to $a=1$ AU, $e=0.8$ have been chosen. Red dashed line: unperturbed Keplerian ellipse at $t = t_0= t_p$. Blue dash-dotted line: osculating Keplerian ellipse after the first perihelion passage. As can be noted, its semimajor axis is larger than that of the initial unperturbed ellipse, while the eccentricity remaines constant. Note also that after 2 yr  the planet has not yet reached the perihelion as it would have done in absence of mass loss.}
\end{figure}
However, it must be noted that a term like \rfr{haj} is inadmissible in any relativistic theory of gravitation because it violates the Lorentz invariance. Indeed, this fact is explicitly shown for general relativity by Bini et al. in \rf{Bin08}  where the full equations of motion of a test particle in a non-stationary gravitoelectromagnetic field are worked out (see, eq. (14) of \rf{Bin08}). In deriving them it is admitted that, in general, $\Phi = \Phi(t, \bds{r})$, but no gravitoelectric terms like \rfr{haj} occur.   Instead, \rfr{accel} is compatible with eq. (14) of \rf{Bin08}.

Schr\"{o}der and Smith in \rf{Sch08}, by assuming the conservation of the angular momentum, derive the orbital expansion by means of equations valid, instead, for orbits with constant radius only, i.e. $v^2/r = \mu(t)/r^2$ and $L=vr$. Then, they assume that not only $v$ but also $r$ vary  and put $v(t)=\sqrt{\mu(t)/r}$, which is, instead, valid for circular orbits of constant radius only, into $L=v(t)r(t) = vr$ getting $\mu(t)r(t)=\mu r$,
where in my notation $r$ and $\mu$ refers to the initial epoch $t_0$. With such an approach  they obtain an expanded terrestrial orbit up to about $2$ times larger than mine.

Noerdlinger in \rf{Nor08}, following Jeans \cite{Jea29} and Kevorkian and Cole \cite{Kev96}, assumes for the variation of a quantity identified by him with the semimajor axis the following expression
\eqI a(t)\mu(t)=a\mu:\lb{jea}\eqF thus, his semimajor axis gets larger. Note that such an equation is the same obtained by \rf{Sch08}. By assuming a variation of $\mu$ linear in time \rfr{jea} would yield an increase of $a$ according to \eqI \dot a = -\left(\rp{\dot\mu}{\mu}\right)a >0;\lb{cazzata}\eqF cfr. with my \rfr{arate}. As a consequence of the constancy of $L^2 = \mu(t) a(t)[1-e(t)^2]$ and of \rfr{jea} he obtains that the eccentricity remains constant, i.e.
\eqI \dot e = 0;\eqF cfr. with my \rfr{erate}. Moreover, another consequence of \rfr{jea} obtained by Noerdlinger in \rf{Nor08} is that the Keplerian period increases as \eqI \rp{P^{\rm Kep}(t)}{P^{\rm Kep}} =\left[\rp{\mu}{\mu(t)}\right]^2;\eqF cfr. with my \rfr{pke}.   Should the quantities dealt with by Noerdlinger are to be identified with the usual osculating Keplerian elements, his results would be incompatible with the real dynamics of  a test particle in the field of a linearly mass-losing body, as I have shown. The quantity obtained by us which exhibits the closest resemblance with \rfr{cazzata}  seems to be the secular variation  of $\Delta r(2\pi)$ for $e=0$. Apart from matters of interpretation, the quantitative results are different. Indeed, I obtain for the Earth a secular variation of the semimajor axis of $-2\times 10^{-4}$ m yr$^{-1}$ and a shift in the radial position along the fixed line of the apsides of $+1.3\times 10^{-2}$ m yr$^{-1}$, while Noerdlinger in \rf{Nor08} gets a secular rate of his semimajor axis, identified with the Astronomical Unit, of about $+1\times 10^{-2}$ m yr$^{-1}$. Note that Noerdlinger uses for the Sun $\dot M/M = -9\times 10^{-14}$ yr$^{-1}$ as in the present work.

Krasinski and Brumberg in \rf{Kra04} deal, among other things, with the problem of a mass-losing Sun  in the framework of the observed secular increase of the Astronomical Unit for which, starting with an equation of motion like \rfr{accel}, they obtain an equation like \rfr{cazzata}. A mass-loss rate of $-3\times 10^{-14}$ yr$^{-1}$, considered somewhat underestimated by Noerdlinger \cite{Nor08}, yields an increase of the Astronomical Unit of  $3\times 10^{-3}$ m yr$^{-1}$. With such a value for $\dot\mu/\mu$ I would obtain a decrease of the semimajor axis of $-7\times 10^{-5}$ m yr$^{-1}$ and an increase in $r$ of $+4\times 10^{-3}$ m yr$^{-1}$.

Concerning the observationally determined increase of the Astronomical Unit, more recent estimates from processing of huge planetary data sets by Pitjeva  point towards a rate of the order of $10^{-2}$ m yr$^{-1}$ \cite{Pit05,Pit08}. It may be noted that my result for the secular variation of the terrestrial radial position on the line of the apsides would agree with such a figure by either assuming a mass loss by the Sun of just $-9\times 10^{-14}$ yr$^{-1}$ or a decrease of the Newtonian gravitational constant $\dot G/G\approx -1\times 10^{-13}$ yr$^{-1}$. Such a value for the temporal variation of $G$ is in agreement with recent upper limits from Lunar Laser Ranging \cite{LLR} $\dot G/G = (2\pm 7)\times 10^{-13}$ yr$^{-1}$.  This possibility is envisaged in \rf{dum} whose authors use
$\dot a/a =-\dot G/G$ by speaking about a small radial
drift of $-(6 \pm 13)\times 10^{-2}$ m yr$^{-1}$ in an orbit at 1 AU.
I, now, apply my analysis to the daily decrement of the semimajor axis of LAGEOS \cite{Ruby} whose relevant orbital parameters are $a=12270$ km, $e=0.0045$. From \rfr{arate}  it turns out that a secular decrease of $a$ of the order of $\approx 1 $ mm d$^{-1}$ could only be  induced by $\dot\mu_{\oplus}/\mu_{\oplus}=-3\times 10^{-6}$ yr$^{-1}$. Clearly, it cannot be due to a variation of the Earth's mass $M$; if it had to be attributed  to a variation of $G$, it would be orders of magnitude too large with respect to the bounds in \rf{LLR} and \rf{dum}. Adopting \rfr{deltaerre} does not substantially alter the situation because the required $\dot\mu_{\oplus}/\mu_{\oplus}$ would be only two orders of magnitude smaller. Now I look at the anomalous increase of the eccentricity of the lunar orbit amounting to $\dot e_{\rm Moon} = (0.9\pm 0.3)\times 10^{-11}$ yr$^{-1}$ \cite{luna}. Such a figure and \rfr{erate}  yield $\dot\mu_{\oplus}/\mu_{\oplus} = 8.5\times 10^{-12}$ yr$^{-1}$; it is one order of magnitude larger than the present-day bounds on $\dot G/G$ \cite{LLR,dum}.
\section{Conclusions}\lb{quattro}
I started in the framework of the two-body Newtonian dynamics by using a radial perturbing acceleration linear in time and straightforwardly treated it with the standard Gaussian scheme. I found that  the osculating semimajor axis $a$, the eccentricity $e$ and the mean anomaly $\mathcal{M}$ secularly decrease while the argument of pericentre $\omega$ remains unchanged; the longitude of the ascending node $\Omega$ and the inclination $i$ are not affected by the phenomenon considered.
%Moreover, the Keplerian period $P^{\rm Kep}$ decreases; it is no longer coincident with the apsidal period $P$ which turns out to be larger than %$P^{\rm Kep}$ and increasing.
The radial distance from the central body, taken on the fixed line of the apsides, experiences a secular increase $\Delta r$. For the Earth, such an effect amounts to about 1.3 cm yr$^{-1}$. By numerically integrating the equations of motion in Cartesian coordinates I found that the real orbital path expands after every revolution, the line of the apsides does not change and the apsidal period is larger than the unperturbed Keplerian one. I have also  clarified that such results are not in contrast with those analytically obtained for the Keplerian orbital elements which, indeed, refer to the osculating ellipses approximating the true trajectory  at each instant.
I also computed the orbital effects of a secular variation of the Sun's mass in the framework of the general relativistic linearized gravitoelectromagnetism  which predicts a perturbing gravitoelectric tangential force proportional to  $\bds{v}/r$. I found that both the semimajor axis and the eccentricity secularly increase; the other Keplerian elements remain constant. Such effects are completely negligible in the present and future evolution of, e.g., the solar system.
%
%I applied our results to the evolution of the Sun-Earth system in the distant future with particular care to the phase in which the Sun, moved to the %RGB of the HR, will expand up to 1.20 AU in order to see if the Earth will avoid to be engulfed by the expanded solar photosphere. Our answer is %negative because, even considering a small acceleration in the process of the solar mass-loss, it turns out that at the end of such a dramatic phase %lasting about $1$ Myr the perihelion distance will have increased by only $\Delta r\approx 0.22-0.25$ AU, contrary to the estimates by \rf{Sch08} %who argue an increment of about $0.37-0.63$ AU. In the case of a circular orbit, the osculating semimajor axis remains unchanged, as confirmed by a %numerical integration of the equations of motion which also shows that the true orbital period increases and is larger than the unperturbed Keplerian %one which remains fixed. Concerning the other planets, while Mercury will be completely engulfed already at the end of the MS, Venus might survive; %however, it should not escape from its fate in the initial phase of the RGB in which the outer planets will experience  increases in the size of their %orbits of the order of $1.2-7.5$ AU.
%%Finally, let us recall that in order to have more accurate results concerning the deep-future evolution of the Earth-Sun system,  also the action of %%the other planets of the solar system should be, in principle, accounted for in numerical simulations including the Sun's mass-loss as well.

As a suggestion to other researchers, it would be very important to complement my analytical two-body calculation by performing  simultaneous long-term numerical integrations of the equations of motion of all the major bodies of the solar system by including a mass-loss term in the dynamical force models as well to see if the N-body interactions in presence of such an effect  may substantially change the picture outlined here. It would be important especially in the RGB phase in which the inner regions of the solar system should dramatically change.

\section*{Acknowledgments}
I thank Prof. K.V. Kholshevnikov, St.Petersburg State University, for useful comments and references.

%-----------------------------------------

\end{document}